\documentclass[float,aps,prc,superscriptaddress,showpacs,twocolumn]{revtex4-1}
\usepackage{bm}
\usepackage{amsmath}
\usepackage{amssymb}
\usepackage{color}
\usepackage{epsfig}
\usepackage[colorlinks,breaklinks]{hyperref}
\usepackage[titletoc]{appendix}
\usepackage{bbm}
\usepackage{multirow}
\newcommand{\be}{\begin{equation}}
\newcommand{\ee}{\end{equation}}

\newcommand{\bes}{\begin{split}}
\newcommand{\ees}{\end{split}}
\newcommand{\ber}{\begin{eqnarray}}
\newcommand{\eer}{\end{eqnarray}}

\newcommand{\bra}{{\langle}}
\newcommand{\ket}{{\rangle}}

\newcommand{\blue}[1]{\textcolor{blue}{#1}}

\newcommand{\hw}{\ensuremath{\hbar\Omega}}
\newcommand{\SU}[1]{\ensuremath{\mathrm{SU}( #1 )}}

\newcommand{\SO}[1]{\ensuremath{\mathrm{SO}( #1 )}}
\newcommand{\On}[1]{\ensuremath{\mathrm{O}( #1 )}}

\newcommand{\SpR}[1]{\ensuremath{\mathrm{Sp}( #1,\mathbb{R} )}}

\usepackage{capt-of}
\usepackage{relsize}
\DeclareMathOperator*{\SumInt}{%
	\mathchoice%
	{\ooalign{$\displaystyle\sum$\cr\hidewidth$\displaystyle\int$\hidewidth\cr}}
	{\ooalign{\raisebox{.14\height}{\scalebox{.7}{$\textstyle\sum$}}\cr\hidewidth$\textstyle\int$\hidewidth\cr}}   
	{\ooalign{\raisebox{.2\height}{\scalebox{.6}{$\scriptstyle\sum$}}\cr$\scriptstyle\int$\cr}}
	{\ooalign{\raisebox{.2\height}{\scalebox{.6}{$\scriptstyle\sum$}}\cr$\scriptstyle\int$\cr}}  
}

\begin{document}

\title {Benchmark calculations of electromagnetic sum rules \\with a symmetry-adapted basis and hyperspherical harmonics}

\author{R.~B. Baker} \affiliation{Department of Physics and Astronomy, Louisiana State University, Baton Rouge, LA 70803, USA}

\author{K.~D. Launey} \affiliation{Department of Physics and Astronomy, Louisiana State University, Baton Rouge, LA 70803, USA}

\author {S. Bacca} \affiliation{Institut f\"ur Kernphysik and PRISMA$^+$ Cluster of Excellence, Johannes Gutenberg-Universit\"at Mainz, 55128 Mainz, Germany} \affiliation{Helmholtz Institute Mainz, (Germany), GSI Helmholtzzentrum f\"ur Schwerionenforschung, Darmstadt, Germany}

\author{N. Nevo Dinur} \affiliation{TRIUMF, 4004 Wesbrook Mall, Vancouver, BC, V6T 2A3, Canada} 

\author{T. Dytrych} \affiliation{Nuclear Physics Institute, Academy of Sciences of the Czech Republic, 25068 Rez, Czech Republic}

\date{March 12, 2020}

\begin{abstract}
We demonstrate the ability to calculate electromagnetic sum rules with the \textit{ab initio} symmetry-adapted no-core shell model. By implementing the Lanczos algorithm, we compute non-energy weighted, energy weighted, and inverse energy weighted sum rules for electric monopole, dipole, and quadrupole transitions in $^4$He using realistic interactions. We benchmark the results with  the hyperspherical harmonics method and show agreement within $2\sigma$, where the uncertainties are estimated from the use of the many-body technique. We investigate the dependence of the results on three different interactions, including chiral potentials, and we report on the $^4$He electric dipole polarizability calculated in the SA-NCSM that reproduces the experimental data and earlier theoretical outcomes. We also detail a novel use of the Lawson procedure to remove the spurious center-of-mass contribution to the sum rules that arises from using laboratory-frame coordinates. We further show that this same technique can be applied in the Lorentz integral transform method, with a view toward studies of electromagnetic reactions for light through medium-mass nuclei.

\end{abstract}


\maketitle

\section{Introduction}
\label{sec:intro}

Electromagnetic transitions in atomic nuclei can reveal important information about the dynamical structure of the nucleus itself.
Due to the perturbative nature of the electromagnetic interaction, calculations of these observables can be compared in a straightforward way to experimental data and important features of the strongly interacting nuclear system can be studied. Considerable progress has been achieved in computing these quantities with \textit{ab initio} methods that describe the nucleus as a system of protons and neutrons interacting with each other as well as with external probes, and solve the problem exactly or with controlled approximations~\cite{Bacca:2014tla}. 
Electromagnetic transitions are calculated as inner products of electromagnetic operators between an initial state, typically the ground state, and excited states. By varying the nuclear excitation energy, one can study the so called response functions, or structure functions, from which electromagnetic cross sections can be computed and compared to experiment.
When excited states are above the break-up threshold, the nucleus breaks into clusters and, depending on the energy, possibly several break-up channels are simultaneously open.
This makes the calculation of response functions and cross sections considerably more complicated (see, e.g., Ref.~\cite{Johnson:2019sps} and references therein).  While it is desirable to compute the full response function, it is sometimes easier to study its energy moments, the so called sum rules, which can be compared to experiment as well. A prominent example is the electric dipole polarizability of a nucleus~\cite{Miorelli2016}, which is the inverse energy weighted sum rule of the dipole response function and for which extensive comparison of \textit{ab initio} calculations to data have been recently performed~\cite{PhysRevC.90.064619,Hagen16,Birkhan:2016qkr,Miorelli2018,Simonis:2019spj}.

Response functions and sum rules have been successfully calculated in the shell model \cite{LuJohnson_PRC2018} or using  \textit{ab initio} methods, such as hyperspherical harmonics (HH) and no-core shell model (NCSM) for light nuclei \cite{Quaglioni2007370, Bacca:2014tla, Stetcu:2007_NPA} or the coupled-cluster (CC) method thus far for closed-shell nuclei \cite{Bacca:2013_PRL, PhysRevC.90.064619, Bacca_2018_JPCS}.
Recent work has illustrated that the reach of \textit{ab initio} methods can now extend into the intermediate- and medium-mass region,
 in particular in terms of structure observables (e.g., Refs.~\cite{BognerHHSBCLR14,LauneyDD16,Jansen2016,Hagen2016, Morris2018,PhysRevC.95.034319,GysbersNatPhys}).
Further, the demonstration that the CC method can examine the closed-shell $^{100}$Sn nucleus \cite{Morris2018} suggests that first principles descriptions, albeit within some approximations, are feasible in heavy nuclei.
This presents a unique opportunity for these methods to investigate the robustness of available nuclear interactions and to study dynamical observables in this heavier mass region.
To this end, the symmetry-adapted no-core shell model
 (SA-NCSM)~\cite{DytrychLMCDVL_PRL12, LauneyDD16, DytrychLDRWRBLB20} 
has been shown to be a valuable approach capable of using only physically-relevant model spaces with dimensions that are only a fraction of the standard NCSM model space, thereby extending the reach of the NCSM toward heavier nuclei while maintaining important physical features, such as collectivity and clustering.

The main purpose of this work is to utilize the Lanczos sum rule method (LSR)~\cite{NND_PRC_2014} and SA-NCSM wave functions to compute sum rules.
This is a first and important step toward first-principle applications to sum rules and reactions for open-shell nuclei up through the medium-mass region.
In this paper we report results for $^4$He, where exact solutions exist in the HH method and allow for a benchmark study using the same realistic nucleon-nucleon (NN) interactions. Several interactions are employed, including chiral potentials, for which the effect of the three-nucleon forces (3NF) is discussed. In addition, the SA-NCSM results calculated in selected model spaces are compared against those in the corresponding complete model spaces, which recover the outcomes of the standard NCSM \cite{NavratilVB00,BarrettNV13}. In these cases, we find good agreement, while using much smaller model spaces, corroborating earlier finding for structure observables and form factors \cite{DytrychHLDMVLO14,DytrychMLDVCLCS11,LauneyDD16}.

Another objective of this paper is to discuss techniques for handling spurious center-of-mass (CM) excitations when using laboratory-frame coordinates to calculate sum rules. Specifically, we detail a novel use of the Lawson procedure to calculate SA-NCSM sum rules, where the CM spuriosity can be removed exactly. This may be generalized for other many-body methods that aim to calculate sum rules using laboratory-frame coordinates.
Finally, we show that the SA-NCSM can be applied to the Lorentz integral transform method (LIT), which can be used to calculate response functions for medium-mass open-shell nuclei.

This paper is organized as follows. In Section \ref{sec:methods} we provide a brief overview of the many-body methods used in the benchmark study. In Section \ref{sec:res}, we present results for $^4$He for various electromagnetic sum rules with different energy weightings using realistic
 interactions. We also discuss the center-of-mass considerations on sum rules and for the LIT. 
Finally, in Section \ref{conc} we present our conclusions.

\section{Theoretical framework}
\label{sec:methods}

\subsection{Symmetry-adapted no-core shell model}
The SA-NCSM framework \cite{LauneyDD16} is an {\it ab initio} no-core shell-model that employs a symmetry-adapted basis.
In this work, we use an \SU{3}-coupled basis. As in the NCSM, the
particle coordinates are specified in the laboratory frame. We employ the many-body $N_{\rm max}$ truncation where we enumerate all many-body states, with the selected symmetries, possessing total harmonic oscillator (HO) excitation quanta less than or equal to $N_{\rm max}$.  Specifically, the $N_{\rm max}$  cutoff is defined as the maximum number of HO quanta allowed in a many-particle state above the minimum for a given nucleus. Hence, basis states where one nucleon carries all the $N_{\max}$ quanta are included, in which cases one nucleon occupies the highest HO shell.

The SA-NCSM allows one to down-select from all possible configurations to a subset that tracks with an inherent preference of a system towards low-spin and high-deformation dominance -- and symplectic multiples thereof in high-$N_{\max}$~spaces \cite{DytrychLMCDVL_PRL12} -- as revealed to be important in realistic NCSM wave functions \cite{DytrychSBDV_PRL07,DytrychSDBV08_review}.

The  many-nucleon basis states of the SA-NCSM are decomposed into spatial and intrinsic spin parts, where the spatial part is further classified according to the \SU{3}$\supset$\SO{3} group chain.  The significance of the \SU{3} group for a
microscopic description of the nuclear collective dynamics can be seen from the fact that it is the symmetry group of the successful Elliott model~\cite{Elliott58,Elliott58b}, and a subgroup of the physically relevant \SpR{3} symplectic model~\cite{RosensteelR77, Rowe85,Rowe96}, which provides a comprehensive theoretical foundation for understanding the dominant symmetries of nuclear collective motion. 

The SA-NCSM basis states are labeled schematically as
\begin{equation} 
|\vec{\gamma}; N(\lambda\,\mu)\kappa L; (S_{p}S_{n})S; J M\rangle,  \label{SAbasis} 
\end{equation}
where $S_{p}$, $S_{n}$, and $S$ denote proton, neutron, and total intrinsic spins, respectively. $N$ is the total number of HO excitation quanta. The values $(\lambda\,\mu)$ represent a set of quantum numbers that labels an \SU{3}
irreducible representation, or  ``irrep'' -- they bring forward important information about nuclear shapes and
deformation, according to an established mapping \cite{CastanosDL88,RosensteelR77,LeschberD87}; for example, $(0 0)$,
$(\lambda\, 0)$ and $(0\,\mu)$ describe spherical, prolate and oblate deformation, respectively. The label $\kappa$ distinguishes multiple occurrences of the same orbital momentum $L$ in the parent irrep $(\lambda\,\mu)$.  The $L$ is coupled with $S$ to the total angular momentum $J$ and its projection $M$.    The symbol $\vec{\gamma}$ schematically denotes the additional quantum numbers needed to specify a distribution of nucleons over the major HO shells and their single-shell and  inter-shell quantum numbers.

The SA-NCSM uses a Hamiltonian that, in its most general form, is given as
\begin{eqnarray}
\hat{H} = \hat{T}_{\mathrm{rel}} + \hat{V}_{\rm NN} + \hat{V}_{\rm 3N} + \hat{V}_{\mathrm{C}} , \label{H}
\end{eqnarray}
where $T_{\mathrm{rel}} = \frac{1}{A}\sum_{i<j} \frac{(\vec{p}_i - \vec{p}_j)^2}{2m}$ is the relative kinetic energy ($m$ is the nucleon mass), $V_{\rm NN (3N)}$ is the nucleon-nucleon (three-nucleon) interaction, and $V_{\mathrm{C}}$ is the Coulomb interaction between the protons.
Similarly to the NCSM, where $N_{\rm max}$ is used to denote the model space, in the SA-NCSM, we adopt a notation where an SA-NCSM model space of ``$\langle N_0 \rangle N_{\max}$'' includes all the basis states up through $N_0$  total excitation quanta and a selected set of basis states in $N_0+2$, $N_0+4$,... up through $N_{\max}$. The selection is based on high-deformation and low-spin dominance, along with symplectic \SpR{3} excitations thereof. Hence, configurations of largest deformation (typically, large $\lambda$ and $\mu$) and lowest spin values are included first. 
This ensures that the SA-NCSM model spaces accommodate highly-deformed configurations with high-energy HO excitations together with essential mixing of low-energy excitations~\cite{DytrychLMCDVL_PRL12,DytrychSBDV_PRL07,DytrychSDBV08_review}.

\subsection{Hyperspherical harmonics}
In the HH method and its effective interaction counterpart~\cite{Barnea:2000_PRC,Barnea:2003_PRC,Barnea:2004_FBS,Barnea:2010_PRC}, the $A$-body problem is solved working in the center-of-mass frame. Starting from particle coordinates, one defines the Jacobi vectors and retains only the ($A-1$) relative vectors, removing the center-of-mass coordinate.
From the relative Jacobi vectors, one then introduces the hyperspherical coordinates, which are constituted by a hyperradius $\rho$ and $(3A-4)$-hyperangles, denoted cumulatively by $\hat{\Omega}$~\cite{Barnea:2000_PRC,Barnea:2003_PRC}. 
The HH wave function is cast into spatial and spin-isospin part (similarly to the SA-NCSM wave functions, which however are given in the proton-neutron formalism).
The spatial part, described by the coordinates $(\rho,\hat{\Omega})$,  is expanded in terms of a product of hyperradial basis states and  hyperspherical harmonics. 
Omitting the isospin for simplicity, the overall basis states are labeled systematically as
\begin{equation}
|n, [K]_A; (S_{p}S_{n})S; J M\rangle,  \label{HHbasis} 
\end{equation}
where $n$ is a hyperradial quantum number -- for example, the order of Laguerre polynomials used to expand the hyperradial wave function -- and
 $[K]_A$ represents a cumulative quantum
number that includes the grandangular momentum quantum number $K$, as well as the angular momentum $L$, while the lower index indicates that the state is antisymmetrized. The antisymmetrization is performed with a powerful algorithm that exploits the group chain \On{3A-3} $\supset$ \On{3} $\otimes$ \On{A-1} $\supset \mathrm{S}_A \otimes$ \On{3} ~\cite{BN,BN2}.

The intrinsic Hamiltonian used is the same as in Eq.~(\ref{H}), where the relative kinetic energy  can be written in hyperspherical coordinates as  
\begin{equation}
\hat{T}_{\rm rel} = \frac{1}{2m}\left[-\varDelta_{\rho} + \frac{\hat{K}^2}{\rho^2}\right].
\end{equation}
Here,  $\varDelta_{\rho}$ only depends on $\rho$ and $\hat{K}^2$ is the hyperangular momentum operator. The latter can be viewed as a generalization of the angular momentum in a multidimensional space.
Because the HH are eigenfunctions of $\hat{K}^2$, with eigenvalues related to hyperspherical quantum number $K$, the relative kinetic energy is diagonal in this basis.
For the general Hamiltonian 
(\ref{H}), the Hamiltonian matrix  on the basis of Eq.~(\ref{HHbasis}) needs to be diagonalized.  In practice, the model space is truncated at some maximal value $K_{\rm max}$ and $n_{\rm max}$
 of the quantum numbers $K$ and $n$, respectively, and  
convergence is reached when  $K_{\rm max}$ and $n_{\rm max}$ are large enough that the calculated observables are independent of these cutoffs~\cite{Barnea:2000_PRC,Barnea:2003_PRC,Barnea:2010_PRC,Goerke:2012ge}.

\subsection{Lanczos sum rule and Lorentz integral transform methods}

The response of a nucleus
to an external perturbation of energy $E_x$ is described 
by the response function, defined as
\begin{equation}\label{eq:S_omega}
R(E_x)=\SumInt_f |\bra \psi_f |\hat{O}| \psi_0 \ket|^2 \delta\left(E_f-E_0-E_x\right),
\end{equation}
where $\hat{O}$ is the operator that induces a transition from the initial state $| \psi_0\ket$ into a set of final states $ |\psi_f \ket$. Here, $| \psi_{0(f)}\ket$ and $E_{0(f)}$ are eigenstates and the corresponding eigenvalues, respectively, of the Hamiltonian $\hat{H}$, 
and $\SumInt_f$ includes the entire discrete and continuous spectrum, such that $\SumInt_f |\psi_f \ket \bra \psi_f | = \mathbf{1}$.

In this work, we focus on several moments of the response function, i.e., sum rules of the form
\begin{equation}\label{eq:SR_mom_integral}
m_n = \int \, dE_x \, R(E_x) \, E_x^n,
\end{equation}
which, using the completeness of the eigenstates $|\psi_f \ket$, can be rewritten as
\begin{equation}\label{eq:SR_mom_braket}
m_n = \bra \psi_0 | \hat{O}^\dagger\, \left( \hat H - E_0\right)^n\, \hat{O} | \psi_0\ket.
\end{equation}
This suggests that the calculation of $m_n$ does not require explicit knowledge of the response function. Furthermore, if the initial state $|\psi_0 \ket$
is localized
and well described within the range of the interaction, 
 then it is justified to use a bound-state method to calculate the wave function $| \psi_0\ket$ and $m_n$ \cite{NND_PRC_2014}.

Of particular interest is the zeroth moment $m_0$ or the square of the norm of the transitional state $\hat{O}|\psi_0 \ket$ 
\begin{equation}\label{eq:Norm}
m_0 = \bra \psi_0 | \hat{O}^\dagger\, \hat{O}|\psi_0 \ket = \int  dE_x R(E_x), 
\end{equation}
which is also known as the non-energy weighted sum rule (NEWSR) or the total strength of the response function. In this paper, besides  $m_0$, we also focus on $m_1$ and $m_{-1}$, which are called the energy weighted sum rule (EWSR) and inverse energy weighted sum rule (IEWSR), respectively, and perform a study of the electric monopole, dipole, and quadrupole operators.

To calculate the sum rules, we use the LSR method (see, e.g., \cite{Dagotto:1994_RMP,NND_PRC_2014} and references therein).
 The LSR method uses
\begin{eqnarray} \label{eqn:LSR_eq}
m_n = \bra \psi_0 | \hat{O}^{\dagger} \hat{O}|\psi_0 \ket \sum_{k=0}^{N_L-1} |Q_{k 0}|^2 (E_{x,k})^n,
\end{eqnarray}
where $N_L$ is the number of Lanczos iterations, $Q_{k 0}$ is the matrix that diagonalizes the tridiagonal Lanczos matrix, and $E_{x,k}$ is the excitation energy of the $k$-th state.
The method benefits
from a suitable choice of the Lanczos pivot, the starting point of the iterative tridiagonalization process. In particular, for the pivot we use the normalized transitional state 
\begin{equation}\label{eq:pivot}
|\phi_0\ket = \frac{ \hat{O}|\psi_0 \ket } {\sqrt{m_0} }.
\end{equation}
The LSR method has been shown to be very efficacious \cite{NND_PRC_2014} and has, for example, allowed to reach the required precision in the calculations of  nuclear structure corrections to the Lamb shift of light muonic atoms~\cite{Ji_PRL_2013, NND_PRC_2014, NND:2016_PhysLettB, Ji_2018}.
Furthermore, the method has been recently applied to calculate $m_{-1}$ for the dipole operator within coupled-cluster theory~\cite{Miorelli2016,Hagen16,Miorelli2018}.

Response functions
can be obtained without explicitly solving for the final eigenstates by utilizing integral transform methods. A prominent example is the Lorentz integral transform, which has been well documented in the literature and used to obtain nuclear responses for electromagnetic and weak operators~\cite{Efros:1994_PLB,Efros:2007_JPG}. 
The Lorentz integral transform is defined as
\begin{eqnarray}
\mathcal{L}(\sigma,\Gamma) = \frac{\Gamma}{\pi} \int dE_x \frac{R(E_x)}{(E_x - \sigma)^2 + \Gamma^2},
\end{eqnarray}
where $\sigma$ and $\Gamma$ determine the peak-position and width of the Lorentzian kernel, respectively. It can be shown that 
\begin{equation}\label{eq:LIT_eq0}
 \mathcal{L}(\sigma,\Gamma)= \bra {\psi} | {\psi} \ket\,,
\end{equation}
where $|{\psi} \ket $
is found as a unique solution of 
the so-called LIT equation
\begin{equation}\label{eq:LIT_eq}
\left( \hat{H} - z \right) |{\psi} \ket =\hat{O}|\psi_0 \ket ,
\end{equation}
where $z=E_0 + \sigma + i\Gamma$. From here, $\mathcal{L}(\sigma,\Gamma)$ is determined by the Lanczos coefficients obtained by iterations from the starting pivot of Eq.~(\ref{eq:pivot}), as shown, e.g., in Eq. (3.40) of Ref.~\cite{Efros:2007_JPG}.

\section{Results}
\label{sec:res}
In this work, the aim is to illustrate the ability of the SA-NCSM  to reliably calculate the necessary nuclear states
required as input to the LSR and LIT methods. To achieve this we focus on $^4$He and begin by studying the convergence of results with increasing model space size, for a given SA selection. We compare these results to computations obtained in the HH. In addition, we present the complete-space SA-NCSM results, which coincide with those obtained in the
standard NCSM
for the same $N_{\rm max}$ (hence, labeled as ``NCSM"). Following this discussion, we detail the efficient way we developed to remove the spurious CM contribution to sum rules and LIT when operators that are not translationally invariant are used,
which may be applicable to other many-body methods.

We discuss these aspects within the context of 
three electromagnetic operators, relevant to
nuclear structure, namely, the isoscalar 
electric monopole (carrying angular momentum $L=0$), the electric dipole ($L=1$), and the electric quadrupole $(L=2)$. These are respectively defined as
\begin{eqnarray}
\hat{M} &=& \frac{1}{2} \sum^A_{i=1} r_i^2 \\
\hat{D} &=& \sqrt{\frac{4\pi}{3}} \sum^A_{i=1} e_i r_i Y_{10}(\hat{r}_i) \label{eqn:dip_op} \\
\hat{Q} &=& \sqrt{\frac{16\pi}{5}} \sum^A_{i=1} e_i r_i^2 Y_{20}(\hat{r}_i),
\end{eqnarray}
where $e_i$ and $\vec r_i$ denote the charge and coordinates of the $i$th particle.  These coordinates can be defined with respect to the center-of-mass, as done in HH that uses translationally invariant operators. In the no-core shell-model framework, $\vec r_i$ are particle coordinates in the laboratory frame, and hence, the operators in Eq.~(\ref{eqn:dip_op}) are not translationally invariant. Consequently, special care is taken to remove the resulting spurious  CM contribution to the SA-NCSM sum rules presented in Sec. \ref{benchmarks} (for details, see Sec. \ref{results_CM}).

For all calculations presented in this paper,
we use well-established NN interactions: JISP16 \cite{ShirokovMZVW07}, N3LO-EM \cite{EntemM03}, and NNLO$_{\rm opt}$ \cite{Ekstrom13}.
We present 
SA-NCSM calculations obtained with $\hw=25$ MeV (unless otherwise indicated), while a range of $\hw$ values between 22 and 28 MeV 
has been used to allow for extrapolations to the infinite model space and estimate uncertainties. 
These extrapolations are based on the Shanks method \cite{Shanks55, BenderO78} to determine the converged value of an infinite sum. 
In particular, one can use the Shanks transformation ansatz for a quantity $X_{\infty}=\sum_{N=0}^{\infty}x_N$ such that  $X_{N_{\rm max}}=\sum_{N=0}^{N_{\rm max}}x_N$ is given by  $X_{N_{\rm max}}=X_\infty +AQ^{N_{\rm max}}$ for large $N_{\rm max}$, where $0<Q<1$. Typically, for data on a converging trend, it is sufficient to use the last three points to determine the infinite-space value as 
\begin{eqnarray}
X_\infty = \frac{X_{N_{\mathrm{max}+2}}X_{N_{\mathrm{max}-2}} - X_{N_{\mathrm{max}}}^2}{X_{N_{\mathrm{max}+2}} + X_{N_{\mathrm{max}-2}} - 2X_{N_{\mathrm{max}}}} ,
\end{eqnarray}
where $X_\infty$ is the converged value of interest and $X_{N_{\rm max}}$ is the 
value calculated at 
$N_{\mathrm{max}}$. This calculation is performed  for each value of $\hw$ and those extrapolated values are used to estimate the combined theoretical uncertainty $\sigma$ in each quantity. Note that these uncertainties are associated with the many-body SA-NCSM model of relevance to the present benchmark study, and do not reflect uncertainties in the interaction used.
\begin{figure}[th]
\centering
\includegraphics[width=0.45\textwidth]{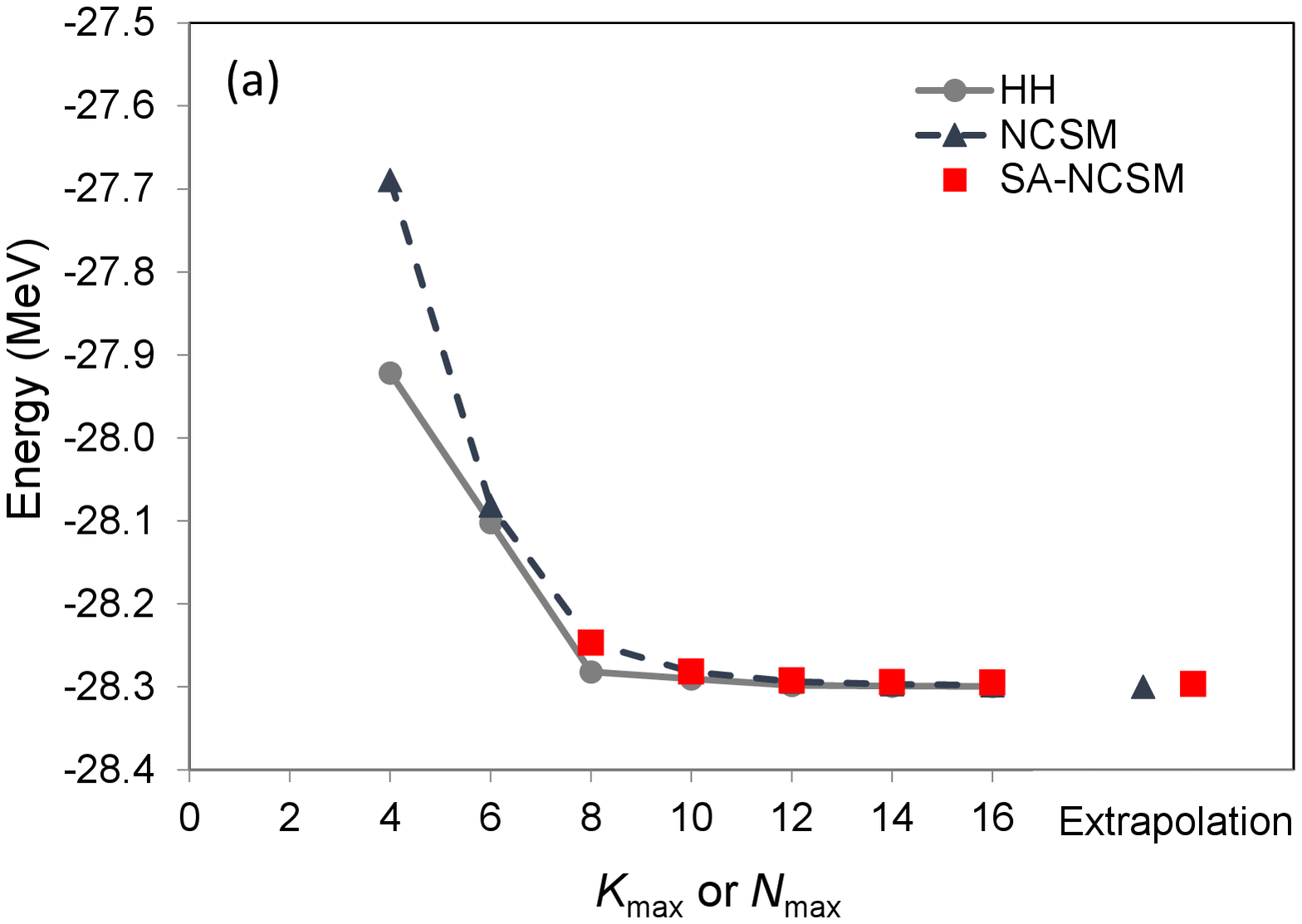} \\
\includegraphics[width=0.45\textwidth]{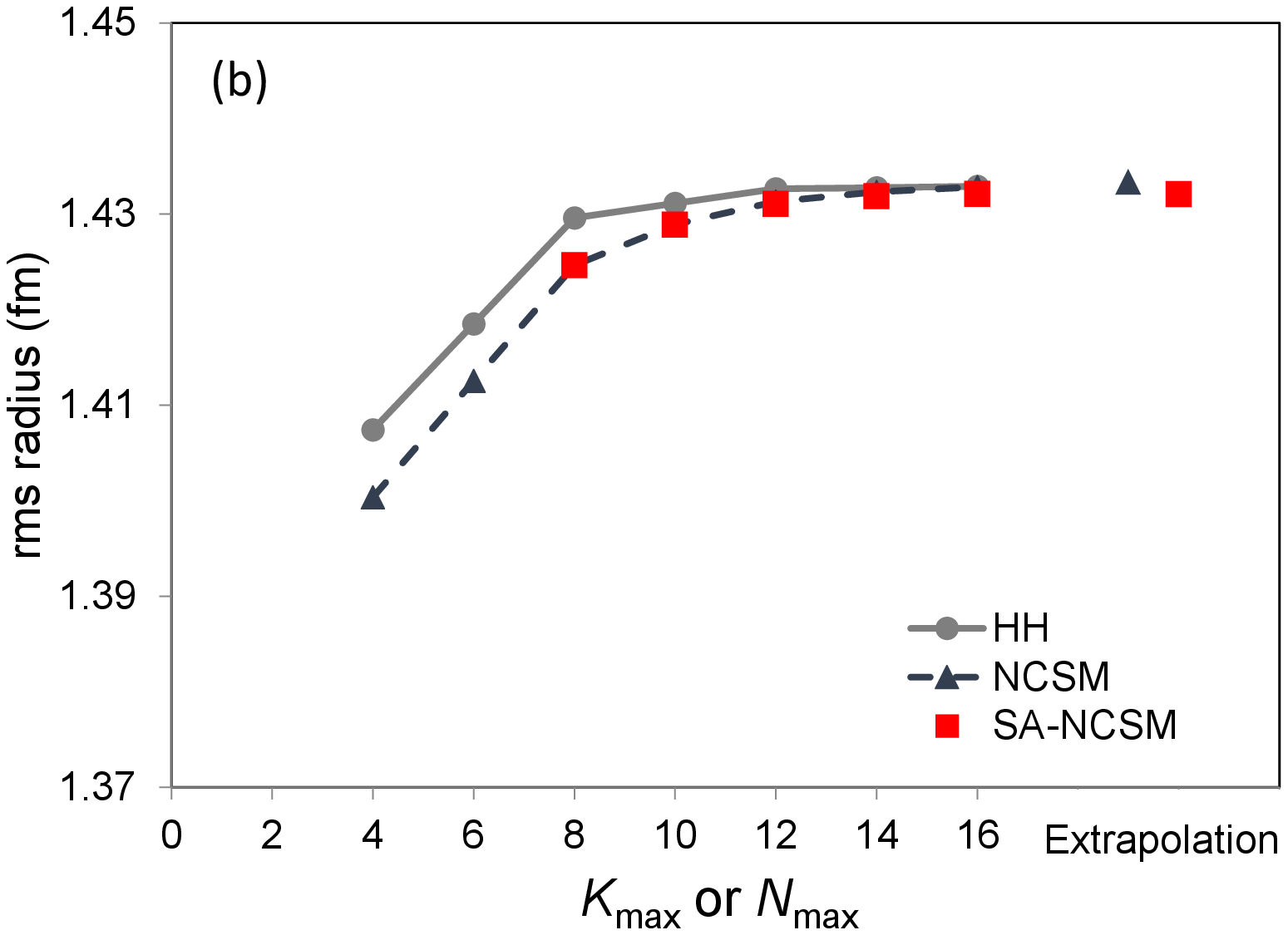}
\caption{Ground-state energy (a) and point-proton rms radius (b) for $^{4}$He as a function of $K_{\mathrm{max}}$ or $N_{\mathrm{max}}$ with the  JISP16 interaction.
NCSM and SA-NCSM points are shown for $\hw=25$ MeV, while the extrapolated values are based on a range of $N_{\mathrm{max}}$ and $\hw$ values.
 Uncertainties of the extrapolated values are smaller than the size of the plot markers.}
\label{gsEnHe4}
\end{figure}

\subsection{Benchmarks for sum rules}
\label{benchmarks}

\begin{table}
\begin{center}
\begin{tabular}{c|lc|lclc}
\hline \hline
					&		\multicolumn{2}{c}{NCSM}				&	\multicolumn{2}{|c}{SA-NCSM}				\\
$J^{\pi}$				&	$N_{\mathrm{max}}$		&	Dimension		&	$N_{\mathrm{max}}$		&	Dimension		\\ \hline
\multirow{3}{*}{$0^+$}	& 	12					&	22,716		&	$\langle 6 \rangle 12$	&	10,357		\\
					&	14					&	58,080		&	$\langle 6 \rangle 14$	&	14,413		\\
					&	16					&	135,475		&	$\langle 6 \rangle 16$	&	14,902		\\ \hline
\multirow{3}{*}{$1^-$}	& 	13					&	103,438		&	$\langle 7 \rangle 13$	&	49,055		\\
					&	15					&	255,074		&	$\langle 7 \rangle 15$	&	56,167		\\
					&	17					&	577,186		&	$\langle 7 \rangle 17$	&	57,547		\\ \hline
\multirow{3}{*}{$2^+$}	& 	12					&	92,958		&	$\langle 6 \rangle 12$	&	31,728		\\
					&	14					&	246,708		&	$\langle 6 \rangle 14$	&	42,226		\\
					&	16					&	591,548		&	$\langle 6 \rangle 16$	&	43,123		\\		
\hline \hline
\end{tabular}
\end{center}
\caption{Dimension of the NCSM and SA-NCSM model spaces for $^4$He for the  relevant 
$J^{\pi}$ states and largest $N_{\mathrm{max}}$ spaces used. The SA-NCSM model spaces are reported in the $\langle N_0 \rangle N_{\max}$ notation (see text for details).
}
\label{tab:dim}
\end{table}
\begin{figure}[th]
\centering
\includegraphics[width=0.45\textwidth]{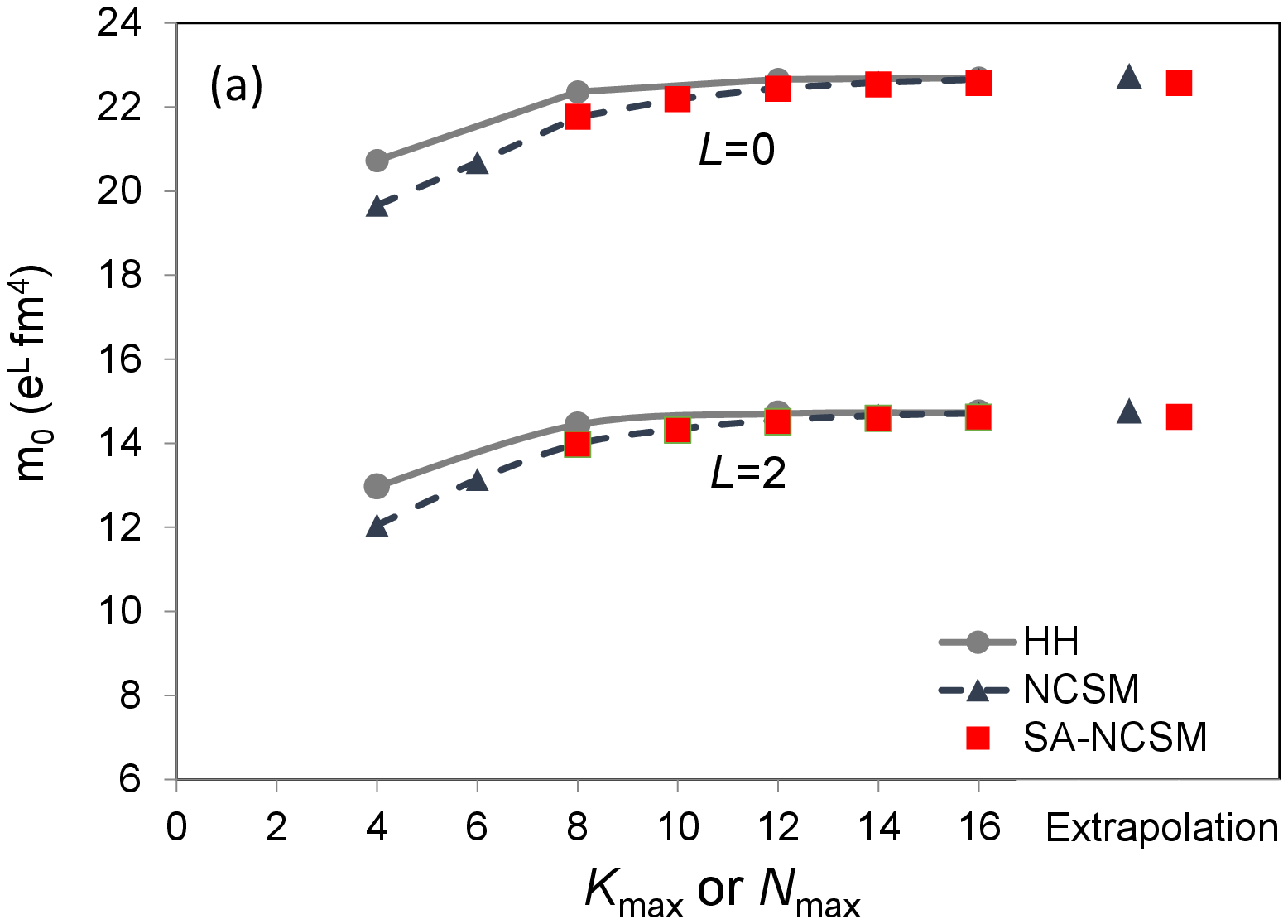}\\
 \includegraphics[width=0.45\textwidth]{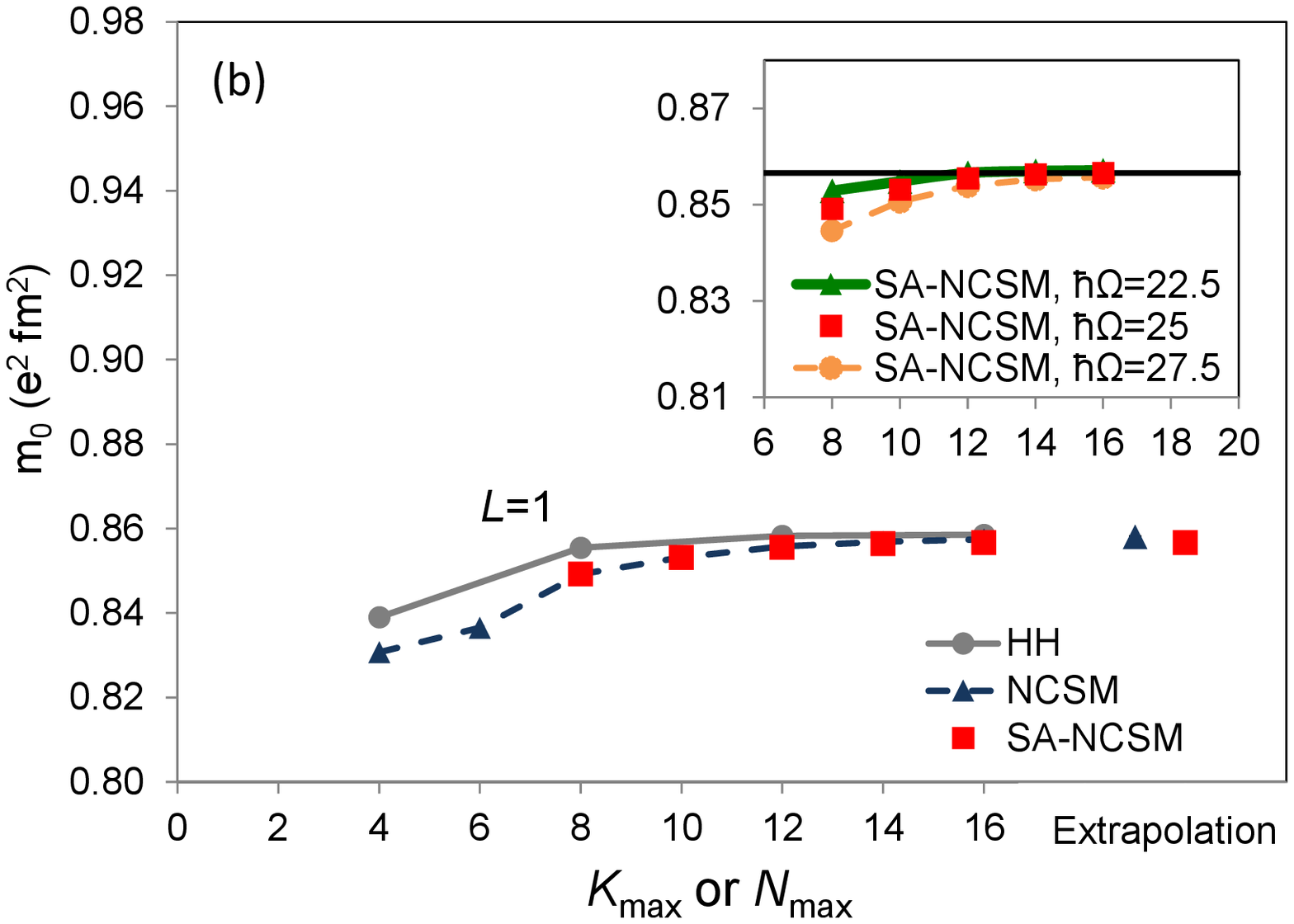}
\caption{Non-energy weighted sum rule as a function of $N_{\mathrm{max}}$ or $K_{\mathrm{max}}$ for $^4$He: (a) total monopole strength ($L=0$) and quadrupole strength ($L=2$), along with (b) dipole strength ($L=1$) and inset showing convergence of three $\hw$ values toward the extrapolated infinite-space value (black solid line). HH, NCSM, and SA-NCSM calculations are performed for the JISP16 interaction; NCSM and SA-NCSM points are shown for $\hw=25$ MeV, while the extrapolated values are based on a range of $N_{\mathrm{max}}$ and $\hw$ values. Uncertainties of the extrapolated values are smaller than the size of the plot markers.}
\label{OLLOvsNmax}
\end{figure}

We start by reporting on the $^4$He ground-state properties, because the 
sum rules for transitions to the ground state 
depend
on the structure of 
the ground-state wave function, in accordance with Eq.~(\ref{eq:SR_mom_braket}).
We present benchmark calculations of 
the ground-state properties of $^4$He within the SA-NCSM and the HH
approaches. We show that both approaches are well converged and agree with each other (Fig.~\ref{gsEnHe4}). In particular,
the ground-state energy and the point-proton rms radius calculated using the JISP16 potential show a relatively quick convergence with 
the model-space size, parameterized by $K_{\rm max}$ for the HH and  $N_{\rm max}$ for the SA-NCSM. This yields small uncertainties for the binding energy and radius when they are extrapolated to their infinite-space values (also shown in Fig.~\ref{gsEnHe4}), which practically coincide with the HH results.
Further, the SA-NCSM using SA selected model spaces is able to reproduce the corresponding complete-space results, or equivalently the NCSM
results, for each $N_{\rm max}$ and for the extrapolated value. This is achieved with a fraction of the basis states used in the SA-NCSM (for model space dimensions, see Table \ref{tab:dim}), while preserving the accuracy of the results, as clearly evident
in Fig.~\ref{gsEnHe4}.

For calculations of sum rules (with no CM spurious contributions),
we find very good convergence with respect to $K_{\mathrm{max}}$ or $N_{\mathrm{max}}$ and agreement between
the SA-NCSM, 
NCSM, and HH models,  as illustrated in Fig.~\ref{OLLOvsNmax} for
$m_0$ for the electric monopole, dipole, and quadrupole operators.
For large $N_{\rm max}$, the SA-NCSM calculations only slightly depend on the \hw~ parameter  (see the inset of Fig.~\ref{OLLOvsNmax}b for a 10\% variation in \hw~around \hw = 25 MeV).
A full comparison of $m_0$, $m_1$ and $m_{-1}$ for the JISP16 interaction is shown in Table~\ref{sumrules_JISP16}. There, we find good overall agreement  within $2\sigma$ 
between the HH results and the extrapolated values for NCSM and SA-NCSM.
In Table \ref{sumrules_JISP16}, we report  $m_1/m_0$ and $m_{-1}/m_0$ 
to avoid compounding the uncertainty of $m_{0}$, as all other sum rules are multiplied by $m_0$ at the end, according to Eq.~(\ref{eqn:LSR_eq}). We note that SA-NCSM monopole and quadrupole $m_{-1}$ moments yield the largest relative uncertainties, which, however, given the very small $m_{-1}$ values, may be numerical in nature and further improved.

\begin{figure}[th]
\centering
\includegraphics[width=0.45\textwidth]{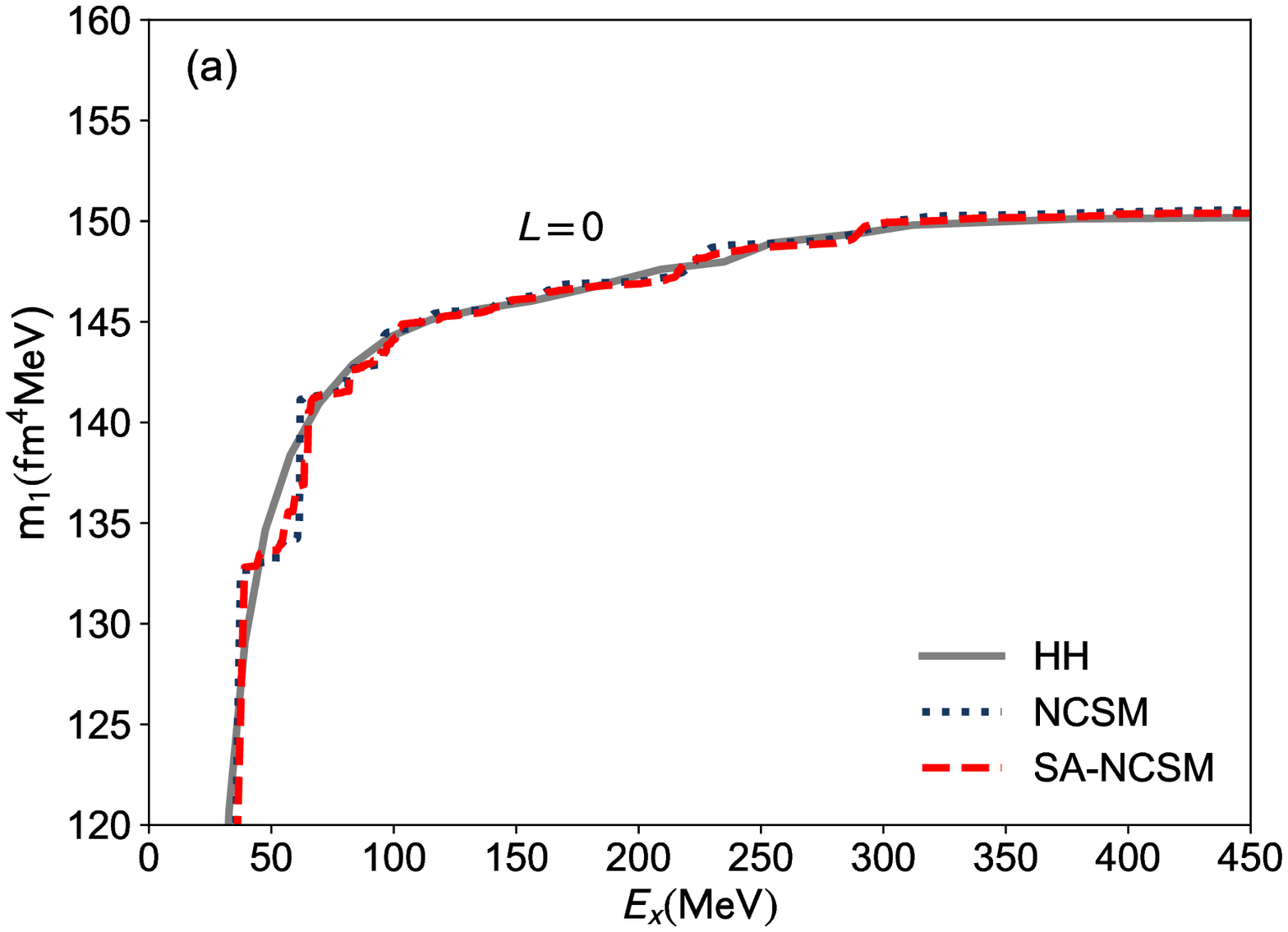} \\
\includegraphics[width=0.45\textwidth]{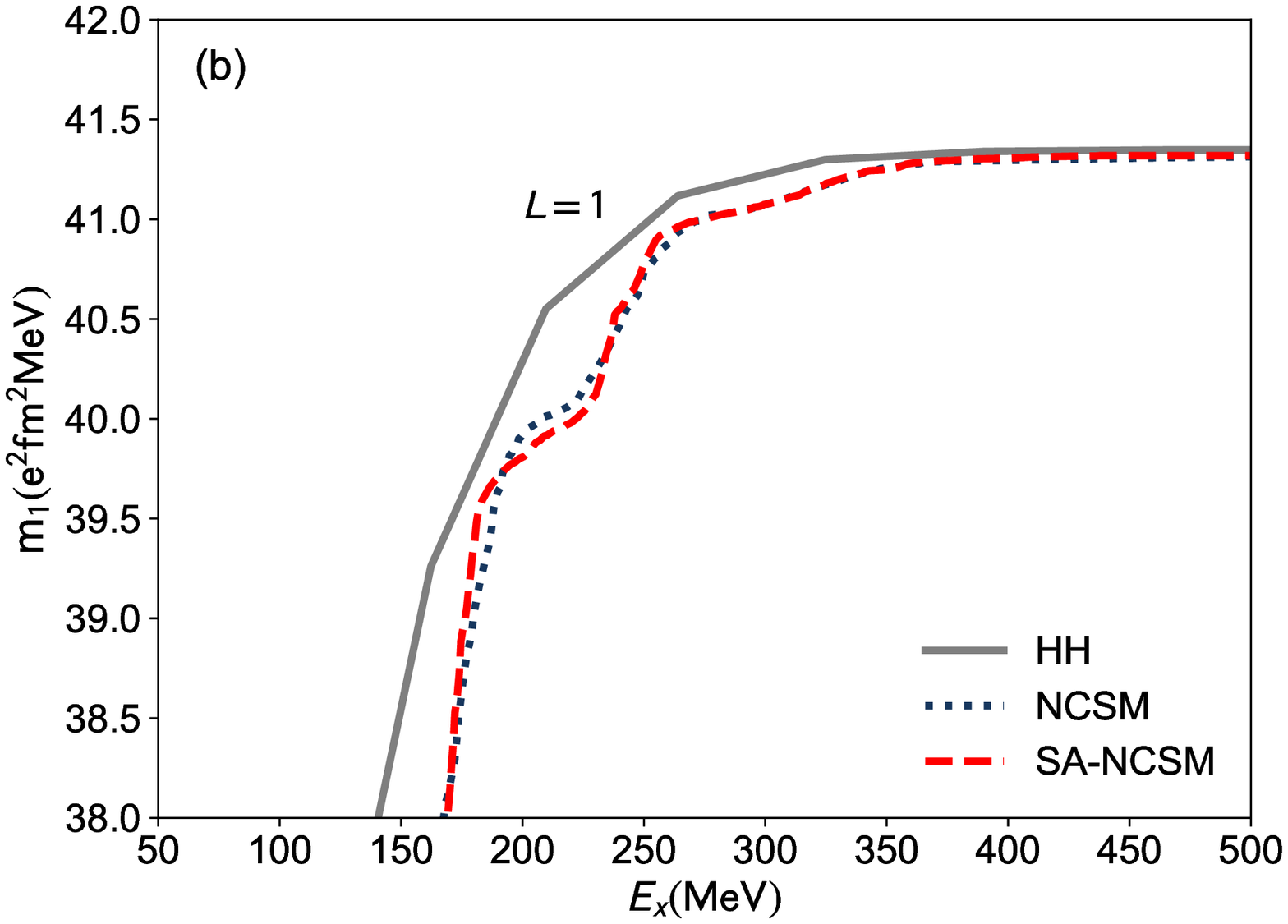}
\caption{Energy weighted sum rules for (a) monopole and (b) dipole transitions as a function of excitation energy for $^4$He. HH, NCSM, and SA-NCSM calculations are performed for the JISP16 interaction; HH results are shown for $K_{\mathrm{max}}=20$ whereas NCSM and SA-NCSM results are shown for $N_{\mathrm{max}}=16$ and $\bra 6 \ket 16$, respectively, with $\hw=25$ MeV.}
\label{fig:EWSR}
\end{figure}

We also examine
the sum rules as a function of the excitation energy, often referred to as running sum rules.
We compare the SA-NCSM and HH calculations for the monopole and dipole energy weighted
running sum rules for the JISP16 interaction (Fig.~\ref{fig:EWSR}). We note that
the detailed structure of these running sums are different. The SA-NCSM and NCSM curves in Fig.~\ref{fig:EWSR} show more discrete jumps, suggesting isolated excited states with some transition strength to the ground state, while the HH curve is smoother due to the higher density of states. 
This fact indicates that the fine details of the excitation spectrum calculated in a discretized basis would be slightly different. 
However and most importantly, as expected from Table~\ref{sumrules_JISP16}, when the sum rules are exhausted by including states at sufficiently large energy, 
the different methods agree and are able to compute converged sum rules with similar accuracy, regardless of the basis used.

\begin{table}[th]
\begin{center}
\begin{tabular}{lccc}
\hline \hline
							&	\multicolumn{3}{c}{JISP16} 								\\ \hline
							&	HH			&	NCSM			&	SA-NCSM			\\
							&						\multicolumn{3}{c}{monopole $L=0$}		\\ \hline
$m_0$ (fm$^4$)				&	22.68(1)		&	22.74(1)			&	22.57(8)			\\
$m_1/m_0$ (MeV)			&	6.623(5)		&	6.63(1)			&	6.67(3)			\\
$m_{-1}/m_0$ (MeV$^{-1}$)	&	0.01103(1)		&	0.01110(1)			&	0.0106(2)			\\
							&						\multicolumn{3}{c}{dipole $L=1$}		\\ \hline
$m_0$ ($e^2$fm$^2$)			&	0.8583(1)		&	0.8581(1)			&	0.8566(7)			\\
$m_1/m_0$ (MeV)			&	48.179(9)		&	48.147(6)			&	48.24(4)			\\
$m_{-1}/m_0$ (MeV$^{-1}$)	&	0.02655(1)		&	0.026574(9)		&	0.02644(8)			\\
							&						\multicolumn{3}{c}{quadrupole $L=2$}	\\ \hline
$m_0$ ($e^2$fm$^4$)			&	14.731(3)		&	14.78(1)			&	14.62(6)			\\
$m_1/m_0$ (MeV)			&	48.98(1)		&	48.92(4)			&	48.5(2)			\\
$m_{-1}/m_0$ (MeV$^{-1}$)	&	0.02543(1)		&	0.02546(3)			&	0.0249(3)			\\
\hline \hline
\end{tabular}
\end{center}
\caption{Non-energy weighted ($m_0$), energy weighted ($m_1$), and inverse energy weighted ($m_{-1}$) sum rules  for monopole, dipole, and quadrupole transitions in $^4$He.  HH, NCSM, and SA-NCSM calculations are performed for the JISP16 interaction. NCSM and SA-NCSM results are the extrapolated values and include estimated uncertainties $\sigma$ based on small variations in $\hw$; uncertainties for HH are obtained  by examining the $K_{\rm max}$ convergence. }
\label{sumrules_JISP16}
\end{table}%

\subsection{SA-NCSM sum rules with chiral potentials}
\begin{figure}[h]
\centering
\includegraphics[width=0.45\textwidth]{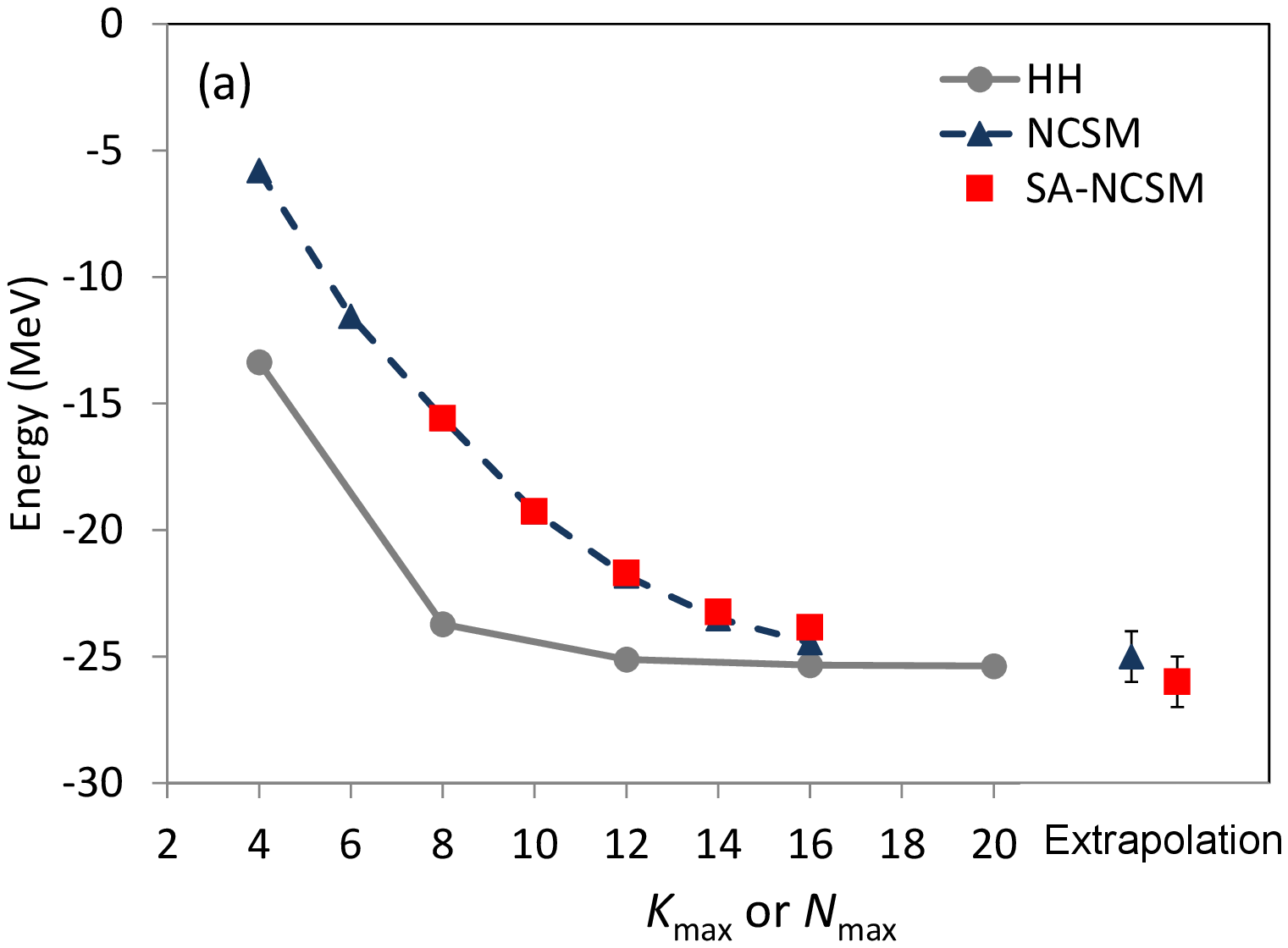} \\
\includegraphics[width=0.45\textwidth]{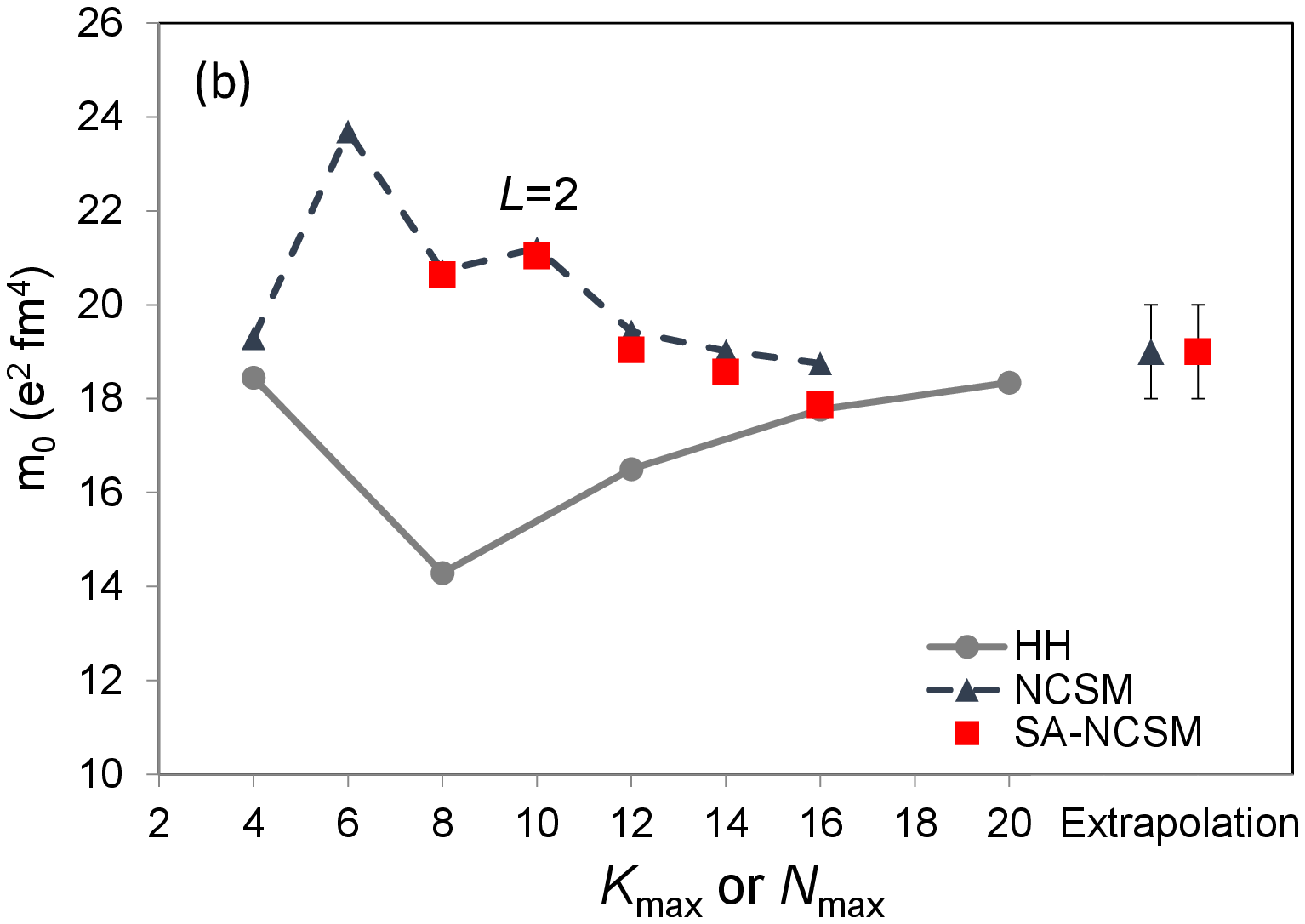}
\caption{(a) Ground-state energy and (b) quadrupole $m_0$ for $^4$He using the N3LO-EM interaction (NN only) as a function of $K_{\mathrm{max}}$ or $N_{\mathrm{max}}$. NCSM and SA-NCSM results are shown for $\hw=25$ MeV,
while the extrapolated values are based on a range of $N_{\mathrm{max}}$ (see Table \ref{tab:dim}) and a 10\% variation in the $\hw$ parameter.
}
\label{fig:N3LO}
\end{figure}

With the goal to explore the dependence of the SA-NCSM results on the nuclear interaction used, we also employ potentials, derived in chiral effective field theory, such as  N3LO-EM \cite{EntemM03} and NNLO$_{\rm opt}$ \cite{Ekstrom13}. To facilitate the comparison, the SA-NCSM calculations consider NN forces only.
As the N3LO-EM is
known not to be as soft as the JISP16, it is interesting to study its convergence properties
and to
perform one last benchmark with the HH method for
the ground-state energy of $^4$He and for $m_0$  (Fig.~\ref{fig:N3LO}). Indeed,
compared to the JISP16, the convergence  for the N3LO-EM is slower for the present approaches,  with the HH method showing a faster convergence rate. For the SA-NCSM and NCSM results, the ground-state energy nears convergence and achieves good agreement with the one calculated in the HH around $N_{\mathrm{max}}=16$ (Fig.~\ref{fig:N3LO}a). We note that the small deviation observed at $N_{\mathrm{max}}=16$ between SA-NCSM and NCSM is a result from the smallest possible SA selection adopted here to illustrate the limits of the SA-NCSM validity. Nevertheless, as for JISP16, 
the extrapolated values 
for N3LO-EM agree remarkably well with the results of the HH within the estimated uncertainties.

Further, it is interesting to point out that for the quadruple  $m_0$ calculated with 
the N3LO-EM interaction (Fig.~\ref{fig:N3LO}b),
the convergence patterns are different between HH and SA-NCSM/NCSM, namely, the first approaching convergence from below and the other from above, whereas convergence rates are comparable.
Here again, the extrapolated  results show a very good  agreement among the models within the respective uncertainties.

\begin{table}[th]
\begin{center}
\begin{tabular}{lcccc}
\hline \hline
						&	\multicolumn{2}{c}{N3LO-EM}		&	\multicolumn{2}{c}{NNLO$_{\mathrm{opt}}$}	\\ \hline
						&	NCSM	&	SA-NCSM		&	NCSM		&	SA-NCSM		\\
						&					\multicolumn{4}{c}{monopole $L=0$}						\\ \hline
$m_0$ (fm$^4$)			&	29(2)		&	29(2)			&	23.29(7)		&	22.9(2)		\\
$m_1$ (fm$^4$MeV)			&	310(50)	&	310(50)		&	177(2)		&	176(1)		\\
$m_{-1}$ (fm$^4$/MeV)		&	0.30(3)	&	0.26(5)		&	0.27(5)		&	0.25(3)		\\
						&					\multicolumn{4}{c}{dipole $L=1$}						\\ \hline
$m_0$ ($e^2$fm$^2$)		&	0.95(3)	&	0.94(2)		&	0.8394(3)		&	0.837(1)		\\
$m_1$ ($e^2$fm$^2$MeV)	&	47(1)		&	46.2(6)		&	39.88(1)		&	39.87(5)		\\
$m_{-1}$ ($e^2$fm$^2$/MeV)	&	0.029(1)	&	0.0268(9)		&	0.0236(1)		&	0.0236(2)		\\
						&					\multicolumn{4}{c}{quadrupole $L=2$}					\\ \hline
$m_0$ ($e^2$fm$^4$)		&	19(1)		&	19(1)			&	15.45(6)		&	15.1(1)		\\
$m_1$ ($e^2$fm$^4$MeV)	&	850(130)	&	900(110)		&	706.3(2)		&	707.8(2)		\\
$m_{-1}$ ($e^2$fm$^4$/MeV)	&	0.52(5)	&	0.47(5)		&	0.23(3)		&	0.20(3)		\\
\hline \hline
\end{tabular}
\end{center}
\caption{Non-energy weighted ($m_0$), energy weighted ($m_1$), and inverse energy weighted ($m_{-1}$) sum rules for monopole, dipole, and quadrupole transitions in $^4$He. NCSM and SA-NCSM calculations are performed for the  N3LO-EM and NNLO$_{\mathrm{opt}}$ interactions \blue{(NN only)}; NCSM and SA-NCSM results are the extrapolated values and include estimated uncertainties $\sigma$ based on small variations in $\hw$.
}
\label{sumrules_chiral}
\end{table}

For comparison purposes, we tabulate $m_0$, $m_1$, and $m_{-1}$  for the monopole, dipole, and quadrupole operators, as calculated in the SA-NCSM using two chiral potentials, N3LO-EM and NNLO$_{\mathrm{opt}}$ (Table~\ref{sumrules_chiral}).
We observe that results do depend on the NN interaction employed,
with the N3LO-EM providing generally larger values for the sum rules than the NNLO$_{\mathrm{opt}}$ (the latter yields smaller values by about 1-20\% relative to the sum rules for the N3LO-EM NN). At the same time, the results for the NNLO$_{\rm opt}$  consistently agree with their counterparts calculated with the JISP16 interaction (c.f.  $m_0$, $m_1/m_0$, and $m_{-1}/m_0$ in Table \ref{sumrules_JISP16}), except a slight increase (decrease) for the monopole $m_1$ (quadrupole $m_{-1}$) sum rule. 
These findings suggest that the complementary 3N forces, omitted in the calculations, have a nonnegligible effect on the sum rules for the N3LO-EM.
This corroborates earlier findings, namely, the 3N forces for N3LO-EM have been shown to give nonnegligible contributions to binding energies and radii (e.g., see \cite{MarisVN13}), whereas the NNLO$_{\rm opt}$ is known to minimize such 3N contributions in $^3$H and $^{3,4}$He \cite{Ekstrom13}.

\begin{figure}[h] 
\centering
\includegraphics[width=0.5\textwidth]{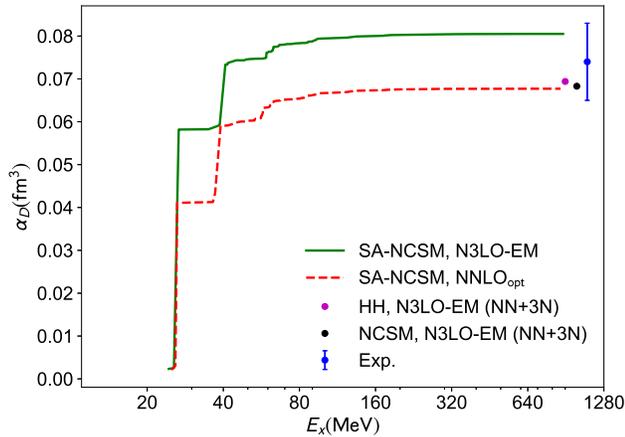}
\caption{Electric dipole polarizability calculated in the SA-NCSM using the N3LO-EM and NNLO$_{\mathrm{opt}}$ interactions (NN only) for $\bra 7 \ket 17$ model spaces and \hw=25 MeV. The values calculated in the HH \cite{Ji_PRL_2013} and the NCSM \cite{PhysRevC.79.064001} with N3LO-EM (N3LO NN $+$ N2LO 3N), together with the experimentally deduced value \cite{Arkatov:1974, Arkatov:1980}, are shown for comparison to the far right of the plot (unrelated to the $E_x$ axis).
}
\label{fig:alphaD}
\end{figure}

To  gain further insight into the properties of the different interactions, we compare the electric dipole 
 polarizability to experiment.
While comparing to data is difficult for the sum rules for the monopole and quadrupole transitions in $^4$He,
in the case of the dipole operator of Eq. (\ref{eqn:dip_op}), the inverse energy weighted sum rule can be related to the electric dipole 
 polarizability $\alpha_D$ as
\begin{eqnarray}
\alpha_D = 2 \alpha \int d E_x \frac{R(E_x)}{E_x} = 2 \alpha\ m_{-1} ,
\end{eqnarray}
where $\alpha$ is the fine-structure constant.
An experimental value for $\alpha_D$ can be extracted from the photoabsoprtion
cross section,
 $\sigma_{\gamma}(E_x)=4\pi^2 \alpha E_x R(E_x)$,  
 by integrating the data~\cite{Arkatov:1974, Arkatov:1980} with the proper energy weight. We show this in Fig.~\ref{fig:alphaD}, along with
 $\alpha_D$ as a function of $E_x$ calculated in the SA-NCSM with the N3LO-EM and NNLO$_{\rm opt}$ NN interactions. Consistent with the outcomes above, the N3LO-EM yields a larger $\alpha_D$ value as compared to the NNLO$_{\rm opt}$, while both results fall within the experimental uncertainties. For further comparison, 
we also include results from previous theoretical work that included the complementary 3N forces in the N3LO-EM,
which has shown that the 3NFs reduce the value of $\alpha_D$ by as much as 15\% \cite{Gazit_PRC_2006}. A remarkable result is that the outcome for the  N3LO-EM (NN$+$3N) closely agrees with that for the NNLO$_{\rm opt}$ using only NN forces, as evident in Fig.~\ref{fig:alphaD}.

\subsection{Treatment of spurious center-of-mass states}
\label{results_CM}

The proper handling of the center-of-mass (CM) excitations is essential for methods that use
laboratory-frame coordinates.
A well-established method to remove CM spuriosity
in the resulting energy spectrum in no-core shell-model calculations
is to use the Lawson procedure ~\cite{Lawson74}  that shifts 
states containing CM excitations to higher energies. This results in low-lying states in the energy region of interest that are translationally invariant.

A very important feature of the SA-NCSM is that any SA-NCSM selected model space
permits exact factorization of the center-of-mass motion of the nuclear system \cite{Verhaar60}. This feature is present in the NCSM, however, it does not hold for any selection of the NCSM model space. In the SA-NCSM, it remains valid only as a result of the SU(3) symmetry used for the selection. Hence, a selected model space yields eigenfunctions that exactly factorize into a product of center-of-mass and intrinsic components, $ |\Psi_{\rm CM} \ket |\psi_{\mathrm{intrinsic}} \ket$. The Lawson procedure~\cite{Lawson74} uses a Lagrange multiplier term that is added to a Hamiltonian expressed in laboratory-frame coordinates, $\hat H + \lambda \hat N_{\mathrm{CM}}$, where $\hat{N}_{\mathrm{CM}}$ is the operator that counts the number of CM excitations and $n_{\mathrm{CM}}$ is its eigenvalue. For a typical value of $\lambda \sim 50$ MeV, the nuclear states of interest (with energy $\lesssim 30$ MeV) have wave functions
that are free of center-of-mass excitations ($n_{\rm CM}=0$), while CM-spurious states ($n_{\rm CM}>0$) lie much higher in energy. 
However, extra care must be taken when calculating observables with these eigenvectors. The reason is that the eigenfunctions are not the intrinsic wave functions, but contain the center-of-mass component with $n_{\rm CM}=0$. Hence,
calculations with operators that are not translationally invariant
can induce CM excitations that affect the results.

 A number of approaches can be used to address this issue. We find two efficient ways: $(i)$ using a CM-free pivot or transitional state $| \phi _0 \ket$ (\ref{eq:pivot}), and $(ii)$ working with a CM-spurious pivot and shifting the CM contribution beyond an energy cutoff, as detailed below. In both cases, to compute the Lanczos coefficients for calculating sum rules and LIT, a Lawson term is used, $\hat H + \lambda \hat N_{\mathrm{CM}}$. Note that this step is in addition to the one that uses the Lawson procedure in the eigenvalue problem to compute the $| \psi _0 \ket$ initial state and that this state is always free of CM excitations. Below we describe both methods in more detail.

\begin{figure}[h] 
\centering
\includegraphics[width=0.5\textwidth]{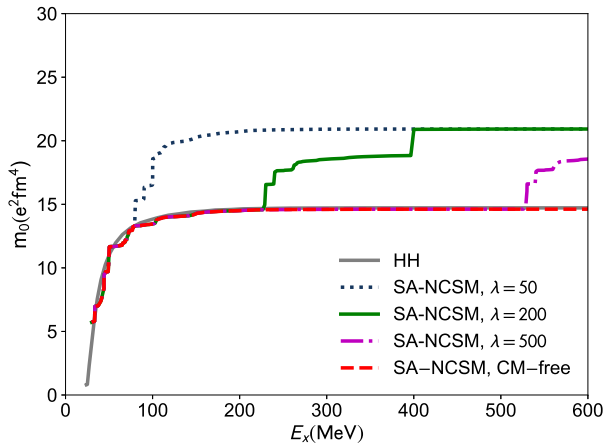} \\
\caption{Quadrupole $m_0$ sum rule computed in the SA-NCSM with the JISP16 in a $\bra 6 \ket 16$ model space for transitions in $^4$He as a function of the excitation energy, calculated by using a CM-free pivot, as well as  a CM-spurious pivot for different values of the Lawson coefficient $\lambda$. The corresponding HH result is also shown for comparison.
}
\label{fig:LSR_CM}
\end{figure}

\noindent
({\it i})  {\it  CM-free pivot. --}
In general, a translationally invariant  transitional state (or pivot) $| \phi _0 \ket$ (\ref{eq:pivot}) can be obtained by using a translationally invariant operator $\hat{O}$, for which the laboratory-frame coordinates
$r_i$, $i=1,2,\dots,A$, are replaced by $(r_i - R_{\rm cm})$ for a center-of-mass coordinate $R_{\rm cm}=\frac{1}{A}\sum_{i} r_i$. This, however, means that one needs to handle many-body operators instead of the original one-body electromagnetic operators. In our work, we adopt an alternative procedure, that is, we project out the CM-free component of the transitional state with  the projection operator 
\begin{eqnarray}
\hat{P} = \prod_{n_{\mathrm{cm}}=1}^{N_{\mathrm{max}}} \left ( \mathbbm{1} - \frac{\hat{N}_{\mathrm{cm}}}{n_{\mathrm{cm}}} \right ).
\end{eqnarray}
This operator selects only the states with $n_{\mathrm{cm}}=0$, thereby removing the contribution of the CM excitations
 up to 
 $N_{\mathrm{max}}$, the model-space cutoff for the $|\psi_i \ket$ 
 wave function. 
The norm can then be calculated, yielding a CM-free $m_0$ moment, which is, in turn, used to calculate the CM-free pivot  
 via Eq.~(\ref{eq:pivot}). The resulting sum rules are CM spuriosity free (see Fig. \ref{fig:LSR_CM}, curve labeled as ``CM-free").

\noindent
({\it ii}) {\it CM-spurious pivot. --}
An alternative approach is to use an operator $\hat{O}$ that is not translationally invariant to obtain a CM-spurious transitional state. The CM-spurious pivot is then calculated using Eq.~(\ref{eq:pivot}), where the CM-spurious norm (or $m_0$) is used. Then
 the normalized pivot vector is used to initiate the Lanczos algorithm for a Hamiltonian that includes a Lawson term, $\lambda \hat N_{\mathrm{CM}}$. This extra term  only acts on CM-spurious states and thus shifts all of them higher in the energy spectrum, as specified by the value of $\lambda$ (see Fig. \ref{fig:LSR_CM}, curves labeled by $\lambda$).
We can then use the Lanczos coefficients in either the LSR or LIT methods. A very important step here is that, for the LSR method, 
we need  to select an energy cutoff to avoid including the higher-lying CM-spurious states,
provided our choice of $\lambda$ is large enough for a given moment $m_n$ to converge.
In Eq.~(\ref{eqn:LSR_eq}), this corresponds to terminating the sum at most at $k_{\rm max} < N_L -1$, such that  $E_{x,k_{\rm max}+1}$ is known to correspond to a CM-spurious state.
Similarly, for the LIT method we can consider an energy range that is appropriate for the response function, provided we have shifted the CM contributions above that region.

To illustrate both procedures, we have calculated the quadrupole $m_0$ sum rule using both the CM-free pivot $(i)$ and the CM-spurious pivot $(ii)$ (Fig.~\ref{fig:LSR_CM}).
The effect of the Lawson term is clearly evident for the CM-spurious pivot: 
we can resolve the contributions to the sum rules from the CM-spurious states that are shifted
above a certain energy specified by the chosen $\lambda$ values of $50, 200,$ and $500$ MeV. Given a large enough $\lambda$, the method can report
a converged value  for the sum rule, provided the convergence is reached at an energy less than $\lambda$. This feature can be found in both NCSM and SA-NCSM calculations.
Similar behaviors have been found for other sum rules and interactions.
Furthermore, the sum rules calculated using 
this technique agree well with those obtained in the HH, where calculations are performed in the center-of-mass frame. They also reproduce the outcome of the SA-NCSM calculations when a CM-free pivot is used.

\begin{figure}[h]
\centering
\includegraphics[width=0.5\textwidth]{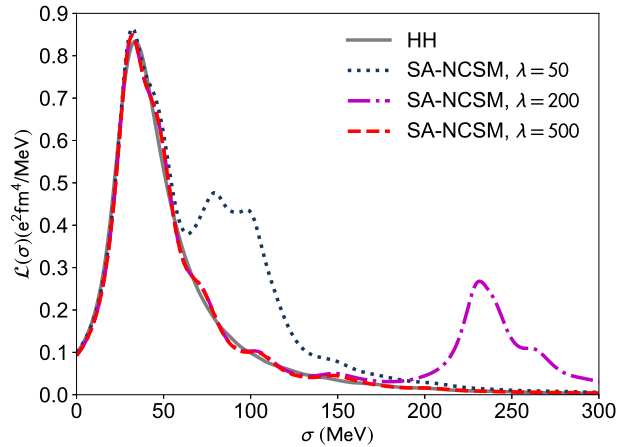}
\caption{LIT with $\Gamma=10$ MeV for quadrupole transitions in $^4$He using the JISP16 potential. Calculations with the SA-NCSM are performed in a $\bra 6 \ket 16$ model space for different values of the Lawson coefficient $\lambda$, and are compared to the HH result.} 
\label{fig:LIT_CM}
\end{figure}

Similar features are observed for the LIT transform, calculated according to Eqs. (\ref{eq:LIT_eq0}) and (\ref{eq:LIT_eq}), as shown in Fig.~\ref{fig:LIT_CM}. Here again, there is a clear evidence of the CM-spurious states, as they shift to higher energies when we increase $\lambda$. Since we must numerically invert the LIT to find the response function, we can use this procedure to shift the CM-spurious contribution above a given energy cutoff and invert the LIT only 
in the energy region below this cutoff. Note that the LIT for any electromagnetic operator depends on the value of the translationally invariant $m_0$, hence, it important to emphasize that the procedure to generate a CM-free LIT transform from a CM-spurious pivot requires two parts: calculate the CM-free $m_0$ (shown in Fig.~\ref{fig:LSR_CM}), and then calculate the LIT curve using the CM-free $m_0$ and a value of $\lambda$ large enough to push the CM-spurious states out of the energy range of interest to 
calculate the response function.
The comparison of the SA-NCSM results for the largest $\lambda$ used and those obtained in the HH  
is very encouraging, and proves that this is a first important step towards studying electromagnetic reactions.

\section{Conclusions}
\label{conc}
We have presented \textit{ab initio} results from the SA-NCSM  for various sum rules describing electric monopole, dipole, and quadrupole transitions in $^4$He, and compared them to those obtained in the HH method. 
We have used the JISP16 and N3LO-EM NN interactions and showed that SA-NCSM calculations reproduce within $2\sigma$ the corresponding
HH outcomes for the same interaction. In addition, the SA-NCSM that uses selected model spaces has been shown to yield results that reproduce the complete-space counterparts, or equivalently,
the NCSM results, albeit with a much smaller model space.
It is important to note that while the fine details of the excitation spectrum calculated in a discretized basis may be slightly different, both HH and SA-NCSM agree and are able to compute converged sum rules with similar accuracy, regardless of the basis used.

To gain further insight into the properties of various realistic interactions, we have calculated the sum rules under consideration in the SA-NCSM using JISP16, N3LO-EM, and 
NNLO$_{\rm opt}$ interactions (NN only). Interestingly, we have found that the JISP16 and NNLO$_{\rm opt}$ yield overall comparable results.
Furthermore, the $\alpha_D$ electric dipole polarizability (related to the inverse energy weighted sum rule) calculated with the NNLO$_{\rm opt}$ closely agrees with the HH and NCSM calculations that use NN+3N (N3LO-EM), as well as to experiment. The difference between the two chiral potentials suggests that the complementary 3N forces of the N3LO-EM may bring forward about 1-20\% reduction in the sum rules, consistent with earlier findings of 15\% in the HH approach.
The benchmark results and comparison to data suggests
that the SA-NCSM can be reliably used to calculate sum rules in light nuclei \cite{BakerSOTANCP42018}.

We have further detailed the use of a new Lawson procedure in the NCSM and SA-NCSM methods to recover translationally invariant sum rules, 
which may have applications in other many-body methods that use laboratory-frame coordinates. We have found that one can use CM-spurious pivot in the Lanczos procedure, by ensuring that a suitable Lawson term is used, that is, a term that shifts the CM-spurious states above an energy cutoff where the sum rules have reached convergence. The sum rules are then reported at this energy cutoff.
Similarly, in the LIT method, which can be used to produce response functions from these methods, a suitable choice for the Lawson term can shift the CM-spurious contribution to energies higher than the region used to invert the LIT transform.

The present outcome lays the foundation that allows us to examine (currently, work in progress) the underlying dynamics of sum rules and response functions for open-shell light- and medium-mass nuclei accessible by the SA-NCSM \cite{BakerPhDThesis2019}.

\begin{acknowledgments}
S.B. and N.N.D. would like to thank Nir Barnea for providing the HH code and for useful discussions. This work was supported in part by the U.S. NSF (OIA-1738287, ACI-1713690, PHY-1913728), SURA, and the Czech SF (16-16772S), and benefitted from high performance computational resources provided by LSU (www.hpc.lsu.edu) and Blue Waters; the Blue Waters sustained-petascale computing project is supported by the National Science Foundation (awards OCI-0725070, ACI-1238993) and the state of Illinois, and is a joint effort of the University of Illinois at Urbana-Champaign and its National Center for Supercomputing Applications. A portion of the computational resources were provided by the National Energy Research Scientific Computing Center and by an INCITE award from the DOE Office of Advanced Scientific Computing. Additional support was provided in part by the Natural Sciences and Engineering Research Council (NSERC), the National Research Council of Canada, by the Deutsche Forschungsgemeinschaft DFG through the Collaborative Research Center [The Low-Energy Frontier of the Standard Model (SFB 1044)], and through the Cluster of Excellence [Precision Physics, Fundamental Interactions and Structure of Matter (PRISMA+ EXC 2118/1)].
\end{acknowledgments}

\bibliographystyle{apsrev4-1}
\bibliography{EIHH,intro,lsu_latest}

\end{document}